\documentclass[12pt]{article}
\usepackage{amsmath,amsfonts,amssymb}

\def\1barra{1\! \hskip -1.1pt {\rm l}}

\def\0barra{{\rm O} \!\hskip -3.7pt {\rm l} }

\title{Adiabatic Approximation in the Density Matrix Approach: Non-Degenerate Systems }

\author{ A.C. Aguiar Pinto$^{(1)}$,  K.M. Fonseca Romero$^{(2)}$, \\
             and    M.T. Thomaz$^{(1)}$
 \\
\\
\baselineskip =10pt
{ \small \it $^{(1)}$ Instituto de F\'\i sica - Univ. Federal  Fluminense
\vspace{-0.2cm}}\\
{\small \it Av. Gal. Milton Tavares de Souza s/n.$\!\!^\circ$,
             \vspace{-0.2cm} }\\
{ \small \it CEP: 24210-340, Niter\'oi, R.J.,   Brazil }  \\
{\small\it $^{(2)}$Universidad Nacional, Facultad de Ciencias,}
\vspace{-0.2cm} \\
{ \small \it   Departamento de  F\'\i sica, Ciudad Universitaria,} 
\vspace{-0.2cm} \\
{ \small \it Bogot\'a, Colombia} \\ }
\date{ }

\begin{document}

\maketitle

\begin{abstract}

We study the adiabatic limit in the density matrix approach for a quantum
 system coupled to a weakly dissipative medium.
The energy spectrum of the quantum model is  supposed to be non-degenerate.
In the absence of dissipation, the geometric phases for
periodic Hamiltonians obtained previously by M.V. Berry are recovered 
in the present approach. We determine the necessary condition satisfied by the
coefficients of the linear expansion of the non-unitary part of the
Liouvillian in order to the imaginary phases acquired by  the
elements of the density matrix, due to  dissipative effects, be  geometric.
The results derived are  model-independent. We apply them to spin $\frac{1}{2}$ model
coupled to reservoir at thermodynamic equilibrium.

\end{abstract}

\vfill

\baselineskip=12pt

\newpage

\baselineskip=18pt

\section{Introduction}

 Since the fascinating work by  Berry in '84\cite{berry} showing 
the existence of geometric phases (path-dependent phases) in 
vector states driven by adiabatic periodic Hamiltonians, authors in the
 literature have looked for geometric phases in other physical contexts.
In particular, Joye {\it et al}.\cite{joye} and Berry\cite{berry2} 
independently showed that the transition probability of instantaneous
eigenstates of non-real Hamiltonians in the non-adiabatic regime gets
an imaginary phase. This imaginary phase was measured by 
Zwanziger {\it et al}. for a two-level system\cite{zwanziger}. The 
appearance of an imaginary correction to the  geometric phase in quantum
models coupled to dissipative media has been discussed
in the literature  \cite{2,2.1,2.2}.  In those references, the non-unitary
evolution of  the quantum system is implemented by a phenomenological
non-hermitian Hamiltonian. This phenomenological approach has been 
extensively applied
to the study of the properties of open quantum systems \cite{baker84,moiseyev98}.

Recently we have considered a spin $\frac{1}{2}$ model in the presence of
an external magnetic field and coupled to a weakly dissipative
medium \cite{JphysA}; this system precesses with constant angular
velocity around a fixed axis. We applied two Lindblad operators to represent
the non-unitary part of the Liouvillians of the quantum system
in contact with two distinct reservoirs. Using the master
equations for these models,
we concluded that the geometric phases acquired by the
spin $\frac{1}{2}$ instantaneous eigenstates of the Hamiltonian were not
modified by the presence of the dissipation. The effective result of the
interaction of the quantum system with the reservoir is
the shrinking of the Bloch vector, which can be used to give a geometric description
to the density matrix \cite{stodolsky,stodolsky1,stodolsky2,Sargent}.

Certainly it is still an open question whether our results in
reference \cite{JphysA} are of general nature or particular to the
dissipative models studied. We remind that they are in opposition to the 
ones derived by the non-hermitian Hamiltonian approach \cite{2,2.1,2.2}.
In order to prove that the results of  \cite{JphysA} are valid in general
for   reservoirs  at equilibrium,
we must have a model-independent approach. We should say that 
master equations are the proper
way to study an open quantum system, 
whereas phenomenological non-hermitian
Hamiltonians are supposed to give a ``{\it bone fide}'' description of 
an open system only when the coherence of states 
is not destroyed by the interaction of the system with
its neighborhood. We point out that the general expression for  the 
imaginary phase acquired by the non-hermitian Hamiltonians is given 
by eqs. (75) and (76) of reference \cite{baker84},
from which we conclude that this imaginary phase has
be geometric for any non-hermitian
Hamiltonian, as derived in references \cite{2}, \cite{2.1} and \cite{2.2} 
for some specific models.
However, by the end of reference \cite{2} Garrison and Wright affirm that
their result should be checked out by a density matrix approach.

In reference \cite{JphysA} the density matrix already appears as our
central object. However, even there our explanation is based on
the time evolution of the instantaneous eigenstates of Hamiltonian.
In the present paper, we rederive the evolution of a quantum system in
the adiabatic approximation directly in the density matrix formulation,
which is the natural approach in the study of quantum
systems coupled to dissipative environment.

The Adiabatic Theorem discussed in references \cite{born,adia_theor,adia_theor1}
applies to quantum systems driven by unitary evolution.
It states that if  initially the system  is in an
eigenstate of the Hamiltonian, its time evolution is fasten, at each time, to an
instantaneous  eigenstate of Hamiltonian  with the same 
original quantum numbers. Therefore, it
is a natural choice to write our density matrix in the basis of the instantaneous
eigenstates of the Hamiltonian and derive its adiabatic limit.

The dynamics of the operator density $\rho(t)$ is given by a Liouville-von
Neumann equation. In the Liouvillian we add a non-unitary term to
take into account the interaction with a dissipative medium, that is

\begin{equation}
\frac{ d  \rho (t) }{d t} = - i [ H(t) , \rho (t) ] + {\cal L}_D \rho (t),
\label{01}
\end{equation}

\noindent where $ H(t)$ is the time-dependent Hamiltonian of the
quantum system and ${\cal L}_D $ is a superoperator that acts on
$\rho (t)$ and is responsible for the non-unitary evolution of the
quantum system.

In the basis of the instantaneous eigenstates
of the Hamiltonian $H(t)$ (i.e., $H(t)|u_i;t\rangle = E_i(t)
|u_i;t\rangle$), the dynamical equations of the elements of the density matrix are

\begin{eqnarray}
\frac{d\rho^H_{ij} (t)}{dt} &=& \sum_l \Big( [\frac{d \langle u_i ; t |}{dt}]| u_l;t \rangle
\rho^H_{lj} (t) + \rho^H_{il} (t) \langle u_l ; t |[\frac{ d | u_j ; t \rangle}{dt}]  \Big)
- \nonumber\\
\nonumber \\
&-& i (E_i (t) - E_j(t) )  \rho^H_{ij} (t) + \langle u_i;t | {\cal L}_D \rho (t) | u_j;t\rangle,
\label{04}
\end{eqnarray}

\noindent where $ \rho^H_{ij} (t) \equiv \langle u_i ; t |
\rho (t)| u_j ; t \rangle$. Not all elements of matrix
$\rho_{ij}^H (t)$ are independent,
due to the constraints that Tr[$\rho (t)$]$=1$ and $\rho_{ij}^H (t)$
be a hermitian matrix.

The time-dependent Hamiltonian that drives the quantum system takes
into account its effective interaction with its neighborhood. This interaction
is realized through a set of time-dependent external parameters
$\vec{\bf R}(t)= ({R}_1(t),
{R}_2(t), \cdots, {R}_m(t))$. By hypothesis, the
time-dependence of Hamiltonian  comes only from the external
parameters ($ H(t)=  H(\vec{\bf R}(t))$). For a periodic
Hamiltonian, there are in general two distinct time scales: the time scale
$T_{ij}$ associated to transitions between two instantaneous
eigenstates $i$ and $j$ of the Hamiltonian (being typically of
the same order of magnitude) and the time scale (period) $T$,
associated to the periodicity of the external parameters
($\vec{\bf R}(T)= \vec{\bf R}(0)$).

Our aim is to  derive  the dynamical equations satisfied by the
elements $\rho_{ij}^H (t)$ in the  limit $T_{ij} \ll T \rightarrow \infty $
\cite{ born, adia_theor,adia_theor1}. We consider a quantum system
with non-degenerate energy spectrum. To present the main details of the
calculation, in section 2 we take  the simplest situation  when the
system is {\bf not} coupled to a dissipative medium, in which
case we have to recover the known results of literature \cite{berry}.
In section 3, this non-degenerate quantum system is coupled to a weakly
dissipative medium  and the density matrix has to be used
to describe the quantum behavior of the system. In
subsection 3.1, we discuss the conditions under which the quantum system
could acquire an imaginary geometric phase due to the presence of dissipation.
In section 4 we apply the results of section 3 to the spin $\frac{1}{2}$ model coupled
to two particular reservoirs at thermodynamic equilibrium to verify the
nature (time-dependent or path-dependent) of their imaginary phases.
 Finally, in section 5, we present our conclusions.


\section{The Quantum Systems in the Absence of \\
                   Dissipation}

In the absence of dissipation, the time evolution of the quantum system in the
adiabatic limit is completely described by its vector state. The adiabatic
evolution of vector states is well done
in references \cite{born,adia_theor,adia_theor1}. The appearance
of the real geometric phase in the unitary evolution of vector states
is beautifully described  in reference \cite{berry}.

Our aim in this section is to consider a simpler physical situation to present the main
details of the calculations of getting the adiabatic limit directly from the matrix density
approach. This have the advantage of checking the correctness of our results.

Eq. (\ref{04}) in the absence of a coupling with a dissipative
medium becomes

\begin{eqnarray}
\frac{d\rho^H_{ij} (t)}{dt} &=& - i\Big[ (E_i(t) - E_j(t)) +
i ( - \langle u_i ; t| \frac{d | u_i ; t \rangle}{dt} +
\langle u_j ; t| \frac{d | u_j ; t \rangle}{dt}) \Big] \rho^H_{ij} (t) -
\nonumber \\
&-& \sum_{l \atop{l\neq i}}  \frac{  \langle u_i ; t | [\frac{d H(t)}{dt}]
| u_l;t \rangle}  {E_l(t) - E_i(t)} \rho^H_{lj} (t) +  \sum_{l \atop{l\neq j}}
\frac{ \langle u_l ; t |[\frac{ d H(t)}{dt} ] | u_j ; t \rangle}{
E_j(t) - E_l(t)} \rho^H_{il}(t).
\label{1}
\end{eqnarray}

Since we intend to get the adiabatic limit of the elements of the
density matrix, we proceed as usual \cite{berry,born,Sun} and
factorize the dynamical phase in
$\rho_{ij}^H (t)$,

\begin{equation}
\tilde{\rho}_{ij}(t) \equiv e^{ i \int_0^t dt^\prime ( E_i (t^\prime) - E_j
(t^\prime))} \rho^H_{ij}(t).
\label{2}
\end{equation}

We distinguish three types of dynamical equations for the
elements $\tilde{\rho}_{ij}(t)$:

\begin{subequations}\label{4-6}

\noindent {\it i}) for the diagonal elements $\tilde{\rho}_{ii}(t)$, where
$i=1,2,\cdots,N-1$, namely,

\begin{eqnarray}
\frac{ d \tilde{\rho}_{ii} (t)}{dt} &=& \sum_{l=1}^{i-1} \frac{1}{E_i(t) - E_l(t)}
2 {\rm Re}\Big[ \langle u_l ; t | [\frac{d H(t)}{dt}] | u_i;t \rangle
e^{i \int_0^t dt^\prime ( E_l (t^\prime) - E_i (t^\prime))} (\tilde{\rho}_{li} (t))^* \Big] +
\nonumber \\
&+& \sum_{l=i+1}^{N} \frac{1}{E_i(t) - E_l(t)}
2 {\rm Re}\Big[ \langle u_l ; t | [\frac{d H(t)}{dt}] | u_i;t \rangle
e^{-i \int_0^t dt^\prime ( E_i (t^\prime) - E_l (t^\prime))} \tilde{\rho}_{il} (t) \Big] ;
\label{4}
\end{eqnarray}


\noindent {\it ii}) for the elements $\tilde{\rho}_{ij}(t)$, where
$i= 1,2,\cdots, N-1$ and $j =i+1, \cdots, N-1$, namely,

\begin{eqnarray}
\frac{d\tilde{\rho}_{ij} (t)}{dt} &=&  \Big[
 - \langle u_i ; t| \frac{d | u_i ; t \rangle}{dt} +
\langle u_j ; t| \frac{d | u_j ; t \rangle}{dt} \Big] \tilde{\rho}_{ij} (t) +
\nonumber \\
\nonumber \\
&+& \sum_{l=1 \atop{l\neq i}}^j  \frac{  \langle u_i ; t | [\frac{d H(t)}{dt}]
| u_l;t \rangle}  {E_i(t) - E_l(t)}
e^{-i \int_0^t dt^\prime ( E_l (t^\prime) - E_i (t^\prime))} \tilde{\rho}_{lj} (t) +
\nonumber \\
\nonumber \\
&+& \sum_{l=j+1}^{N}  \frac{  \langle u_i ; t | [\frac{d H(t)}{dt}]
| u_l;t \rangle}  {E_i(t) - E_l(t)}
e^{i \int_0^t dt^\prime ( E_i (t^\prime) - E_l (t^\prime))} (\tilde{\rho}_{jl} (t))^*  +
\nonumber \\
\nonumber \\
&+& \sum_{l=1}^{i-1}
\frac{ \langle u_l ; t |[\frac{ d H(t)}{dt} ] | u_j ; t \rangle}{
E_j(t) - E_l(t)}
e^{i \int_0^t dt^\prime ( E_l (t^\prime) - E_j (t^\prime))} (\tilde{\rho}_{li}(t))^* +
\nonumber \\
&+& \sum_{l=i \atop{l\neq j}}^{N}
\frac{ \langle u_l ; t |[\frac{ d H(t)}{dt} ] | u_j ; t \rangle}{
E_j(t) - E_l(t)}
e^{-i \int_0^t dt^\prime ( E_j (t^\prime) - E_l (t^\prime))} \tilde{\rho}_{il}(t);
\label{5}
\end{eqnarray}


\noindent {\it iii}) for the elements $\tilde{\rho}_{iN}(t)$, where
$i= 1,2,\cdots, N-1$, namely,

\vspace{-0.5cm}

\begin{eqnarray}
\frac{d\tilde{\rho}_{iN} (t)}{dt} &=& \Big[ - \langle u_i ; t|
\frac{d | u_i ; t \rangle}{dt} -
\langle u_N ; t| \frac{d | u_N ; t \rangle}{dt} \Big] \tilde{\rho}_{iN} (t) +
\nonumber \\
&-& \sum_{l=1}^{N-1}  \frac{  \langle u_i ; t | [\frac{d H(t)}{dt}]
| u_N;t \rangle}  {E_i(t) - E_N(t)}
e^{i \int_0^t dt^\prime ( E_i (t^\prime) - E_N (t^\prime))} \tilde{\rho}_{ll} (t) +
\nonumber \\
&+& \sum_{l=1 \atop{l \neq i}}^{N-1}  \frac{  \langle u_i ; t |
[\frac{d H(t)}{dt}] | u_l;t \rangle}  {E_i(t) - E_l(t)}
e^{-i \int_0^t dt^\prime ( E_l (t^\prime) - E_i (t^\prime))} \tilde{\rho}_{lN} (t)  +
\nonumber \\
&+& \sum_{l=1}^{i-1}
\frac{ \langle u_l ; t |[\frac{ d H(t)}{dt} ] | u_N ; t \rangle}{
E_N(t) - E_l(t)}
e^{i \int_0^t dt^\prime ( E_l (t^\prime) - E_N (t^\prime))} (\tilde{\rho}_{li}(t))^* +
\nonumber \\
&+& \sum_{l=i}^{N-1}
\frac{ \langle u_l ; t |[\frac{ d H(t)}{dt} ] | u_N ; t \rangle}{
E_N(t) - E_l(t)}
e^{-i \int_0^t dt^\prime ( E_N (t^\prime) - E_l (t^\prime))} \tilde{\rho}_{il}(t) +
\nonumber \\
&+& \frac{  \langle u_i ; t | [\frac{d H(t)}{dt}]
| u_N;t \rangle}  {E_i(t) - E_N(t)}
e^{i \int_0^t dt^\prime ( E_i (t^\prime) - E_N (t^\prime))} .
\label{6}
\end{eqnarray}

\end{subequations}

\noindent Eqs. (\ref{4}) and (\ref{5}) can be written as Volterra
integral equations \cite{7} of the first type, whereas eq.
(\ref{6}) can be written as a Volterra integral equation of the
second type.

Following references \cite{born} and \cite{Sun} we redefine the time
scale using the period $T$ of the external parameters

\begin{equation}
s \equiv \frac{t}{T}. \label{7}
\end{equation}

\noindent In this new variable $s$, the  integral equations
obtained from eqs. (\ref{4-6}) are:

\vspace{.2cm}

\begin{subequations} \label{8-10}

\noindent i) for the diagonal elements $\tilde{\rho}_{ii}(s)$, where
$i=1,2,\cdots,N-1$,

\begin{eqnarray}
\tilde{\rho}_{ii} (s) &=& \tilde{\rho}_{ii} (0)  + \nonumber \\
& & \hspace{-3cm} +
\sum_{l=1}^{i-1} \int_0^s ds^\prime
\frac{1}{E_i(s^\prime) - E_l(s^\prime)}
2 {\rm Re}\Big[ \langle u_l ; s^\prime | [\frac{d H(s^\prime)}{ds^\prime}]
| u_i;s^\prime \rangle
e^{i T \int_0^{s^\prime} ds^{\prime\prime} ( E_l (s^{\prime\prime}) -
E_i (s^{\prime\prime}))} (\tilde{\rho}_{li} (s^\prime))^* \Big] +
\nonumber \\
& & \hspace{-3cm} + \sum_{l=i+1}^{N} \int_0^s ds^\prime
\frac{1}{E_i(s^\prime) - E_l(s^\prime)}
2 {\rm Re}\Big[ \langle u_l ; s^\prime | [\frac{d H(s^\prime)}{ds^\prime}]
| u_i;s^\prime \rangle
e^{-i T \int_0^{s^\prime} ds^{\prime\prime} ( E_i (s{^\prime\prime}) -
E_l (s^{\prime\prime}))} \tilde{\rho}_{il} (s^\prime) \Big] ;
\label{8}
\end{eqnarray}


\noindent ii) for the elements $\tilde{\rho}_{ij}(s)$, where
$i= 1,2,\cdots, N-1$ and $j =i+1, \cdots, N-1$,

\begin{eqnarray}
\tilde{\rho}_{ij} (s) &=& \tilde{\rho}_{ij} (0)   +
\nonumber \\
&+& \int_0^s ds^\prime
\Big[ - \langle u_i ; s^\prime| \frac{d | u_i ; s^\prime \rangle}{ds^\prime} +
\langle u_j ; s^\prime| \frac{d | u_j ; s^\prime \rangle}{ds^\prime} \Big]
\tilde{\rho}_{ij} (s^\prime) +
\nonumber \\
&+& \int_0^s ds^\prime
\sum_{l=1 \atop{l\neq i}}^j  \frac{  \langle u_i ; s^\prime |
[\frac{d H(s^\prime)}{ds^\prime}]
| u_l;s^\prime \rangle}  {E_i(s^\prime) - E_l(s^\prime)}
e^{-i T \int_0^{s^\prime} ds^{\prime\prime} ( E_l (s^{\prime\prime}) -
E_i (s^{\prime\prime}))} \tilde{\rho}_{lj} (s^\prime) +
\nonumber \\
&+& \int_0^s ds^\prime
\sum_{l=j+1}^{N}  \frac{  \langle u_i ; s^\prime | [\frac{d H(s^\prime)}{ds^\prime}]
| u_l;s^\prime \rangle}  {E_i(s^\prime) - E_l(s^\prime)}
e^{i T \int_0^{s^\prime} ds^{\prime\prime} ( E_i (s^{\prime\prime}) - E_l (s^{\prime\prime}))}
(\tilde{\rho}_{jl} (s^\prime))^*  +
\nonumber \\
&+& \int_0^s ds^\prime
\sum_{l=1}^{i-1}
\frac{ \langle u_l ; s^\prime |[\frac{ d H(s^\prime)}{ds^\prime} ] | u_j ; s^\prime \rangle}{
E_j(s^\prime) - E_l(s^\prime)}
e^{i T \int_0^{s^\prime} ds^{\prime\prime} ( E_l (s^{\prime\prime}) - E_j (s^{\prime\prime}))}
(\tilde{\rho}_{li}(s^\prime))^* +
\nonumber \\
&+& \int_0^s ds^\prime
\sum_{l=i \atop{l\neq j}}^{N}
\frac{ \langle u_l ; s^\prime |[\frac{ d H(s^\prime)}{ds^\prime} ] | u_j ; s^\prime \rangle}{
E_j(s^\prime) - E_l(s^\prime)}
e^{-i T \int_0^{s^\prime} ds^{\prime\prime} ( E_j (s^{\prime\prime}) - E_l (s^{\prime\prime}))}
\tilde{\rho}_{il}(s^\prime);
\label{9}
\end{eqnarray}

\vspace{.7cm}

\noindent iii) for the elements $\tilde{\rho}_{iN}(s)$, where
$i= 1,2,\cdots, N-1$,

\begin{eqnarray}
\tilde{\rho}_{iN} (s) &=& \tilde{\rho}_{iN} (0) -
\nonumber \\
&-&  \int_0^s ds^\prime \Big[ \langle u_i ; s^\prime|
     \frac{d | u_i ; s^\prime \rangle}{ds^\prime} -
\langle u_N ; s^\prime| \frac{d | u_N ; s^\prime \rangle}{ds^\prime} \Big]
         \tilde{\rho}_{iN} (s^\prime) -
\nonumber \\
&-&  \int_0^s ds^\prime
\sum_{l=1}^{N-1}  \frac{  \langle u_i ; s^\prime | [\frac{d H(s^\prime)}{ds^\prime}]
| u_N;s^\prime \rangle}  {E_i(s^\prime) - E_N(s^\prime)}
e^{i T \int_0^{s^\prime} ds^{\prime\prime} ( E_i (s^{\prime\prime}) - E_N (s^{\prime\prime}))}
\tilde{\rho}_{ll} (s^\prime) +
\nonumber \\
&+& \int_0^s ds^\prime
\sum_{l=1 \atop{l \neq i}}^{N-1}  \frac{  \langle u_i ; s^\prime |
[\frac{d H(s^\prime)}{ds^\prime}] | u_l;s^\prime \rangle}  {E_i(s^\prime) - E_l(s^\prime)}
e^{-i T \int_0^{s^\prime} ds^{\prime\prime} ( E_l (s^{\prime\prime}) - E_i (s^{\prime\prime}))}
\tilde{\rho}_{lN} (s^\prime)  +
\nonumber \\
&+& \int_0^s ds^\prime \sum_{l=1}^{i-1}
\frac{ \langle u_l ; s^\prime |[\frac{ d H(s^\prime)}{ds^\prime} ] | u_N ; s^\prime \rangle}{
E_N(s^\prime) - E_l(s^\prime)}
e^{i T \int_0^{s^\prime} ds^{\prime\prime} ( E_l (s^{\prime\prime}) - E_N (s^{\prime\prime}))}
(\tilde{\rho}_{li}(s^\prime))^* +
\nonumber \\
&+& \int_0^s ds^\prime \sum_{l=i}^{N-1}
\frac{ \langle u_l ; s^\prime |[\frac{ d H(s^\prime)}{ds^\prime} ] | u_N ; s^\prime \rangle}{
E_N(s^\prime) - E_l(s^\prime)}
e^{-i T \int_0^{s^\prime} ds^{\prime\prime} ( E_N (s^{\prime\prime}) - E_l (s^{\prime\prime}))}
\tilde{\rho}_{il}(s^\prime) +
\nonumber \\
&+& \int_0^s ds^\prime \frac{  \langle u_i ; s^\prime | [\frac{d H(s^\prime)}{ds^\prime}]
| u_N;s^\prime \rangle}  {E_i(s^\prime) - E_N(s^\prime)}
e^{i T \int_0^{s^\prime} ds^{\prime\prime} ( E_i (s^{\prime\prime}) - E_N (s^{\prime\prime}))} .
\label{10}
\end{eqnarray}

\end{subequations}

Some of the integrals on the r.h.s. of eqs. (\ref{8-10}) are of the following type

\begin{eqnarray}
I_{il}(s)= \int_0^s ds^\prime
\frac{ \langle u_i ; s^\prime |[\frac{ d H(s^\prime)}{ds^\prime} ] | u_l ; s^\prime \rangle}{
E_i(s^\prime) - E_l(s^\prime)}
e^{ i T \int_0^{s^\prime} ds^{\prime\prime} ( E_i (s^{\prime\prime}) - E_l (s^{\prime\prime}))}
\tilde{\rho}(s^\prime),
\label{11}
\end{eqnarray}

\noindent with $i \neq l$ and $\tilde{\rho}(s^\prime)$ representing
one element of the density matrix.

By defining

\begin{equation}
g_{il}(s) \equiv \int_0^s ds^\prime ( E_i (s^{\prime}) - E_l (s^{\prime})),
\label{12}
\end{equation}

\noindent we recognize $I_{il}(s)$ as the Stieltjes integral \cite{8}

\begin{equation}
I_{il}(s)= \int_0^s ds^\prime F_{il}(s^\prime) e^{i T g_{il}(s^\prime)}
\hspace{1cm} {\rm so} \; {\rm that} \hspace{1cm}
F_{il}(s^\prime) = \frac{ \langle u_i ; s^\prime |[\frac{ d H(s^\prime)}{ds^\prime} ]
| u_l ; s^\prime \rangle}{ \dot{g}_{il}(s^\prime)}
\tilde{\rho}(s^\prime),
\label{13}
\end{equation}

\noindent and $ \dot{g}_{il}(s) = \frac{ dg_{il}(s)}{ds}$ (we are {\bf not}
summing over the indices $i$ and $l$).

Assuming that $ F_{il}(s^\prime)$ is a piece-wise continuous function in the
interval [$0,s$], then the Riemann-Lebesgue Theorem \cite{9} gives the value of
$I_{il}(s)$ in the adiabatic limit ($T\rightarrow \infty$)

\begin{equation}
\lim_{\hspace{-.08cm} T \rightarrow \infty} \int_0^s F_{il}(s^\prime) e^{iTg_{il}(s^\prime)}
dg_{il}(s^\prime) = 0,
\label{14}
\end{equation}

\noindent which means that for large but finite values of $T$, the
r.h.s. of eq. (\ref{13}) can be written in terms of the inverse powers of $T$.
Integrating eq. (\ref{13}) by parts, we get

\begin{eqnarray}
I_{il}(s) &=& \frac{1}{iT} \Big[ \frac{ F_{il}(s)}{\dot{g}_{il}(s)}
e^{iTg_{il}(s)} \Big] |_{s=0}^{s}
- \frac{1}{iT} \int_0^s ds^\prime e^{iTg_{il}(s^\prime)}
\frac{ d (F_{il}(s^\prime)/ \dot{g}_{il}(s^\prime) )}
{ds^\prime} \nonumber \\
&=& \frac{1}{iT} \Big[ \frac{ F_{il}(s)}{\dot{g}_{il}(s)}
e^{iTg_{il}(s)} \Big] \Big|_{s=0}^{s} + {{o}}( \frac{1}{iT} )    .
\label{15}
\end{eqnarray}

From the  result (\ref{15}) we obtain  that the
elements $\tilde{\rho}_{ij} (s)$ can be expanded in
powers of ($\frac{1}{T}$) \cite{Sun},

\begin{equation}
\tilde{\rho}_{ij}(s) = \sum_{n=0}^\infty \Big( \frac{i}{T} \Big)^n
\tilde{\rho}_{ij}^{(n)}(s),
\label{16}
\end{equation}

\noindent where $\tilde{\rho}_{ij}^{(n)}(s)$ is the coefficient of
order ($\frac{1}{T}$)$^n$ in the expansion.

We are interested in the adiabatic limit ($T\rightarrow \infty$). This correspond
to substitute  expansion (\ref{16}) in eqs. (\ref{8-10}) and
keeping only terms of order ($\frac{1}{T}$)$^0$ in the differential
equations. At this order, the equations become:

\vspace{.3cm}

\noindent i) for the diagonal elements $\tilde{\rho}_{ii}(s)$,
$i=1,2,\cdots,N-1$.

\begin{equation}
\frac{d \tilde{\rho}_{ii}(s)}{ds} = 0 \hspace{1cm}
\Rightarrow  \hspace{1cm}  {\rho}^H_{ii}(s) =
{\rho}^H_{ii}(0) .
\label{17}
\end{equation}

\noindent This result gives us the meaning of the Adiabatic Theorem
in the density matrix approach: the population of an instantaneous
eigenstate of the Hamiltonian does not change in an adiabatic
process.

\vspace{.7cm}

\noindent ii) for the elements $\tilde{\rho}_{ij}(s)$, where
$i= 1,2,\cdots, N-1$ and $j =i+1, \cdots, N$.

\begin{equation}
\frac{d\tilde{\rho}^{(0)}_{ij}(s)}{ds} =
\Big[ - \langle u_i ; s| \frac{d | u_i ; s \rangle}{ds} +
\langle u_j ; s| \frac{d | u_j ; s \rangle}{ds} \Big]
\tilde{\rho}^{(0)}_{ij} (s)
\label{18}
\end{equation}

\noindent whose solution in the variable $t$ is

\begin{equation}
{\rho}^H_{ij}(t) = e^{ \int_0^t dt^\prime
\Big[ - \langle u_i ; t^\prime| \frac{d | u_i ; t^\prime \rangle}{dt^\prime} +
\langle u_j ; t^\prime| \frac{d | u_j ; t^\prime \rangle}{dt^\prime} \Big]}
e^{-i \int_0^t dt^\prime ( E_i (t^\prime) - E_j
(t^\prime))} {\rho}^H_{ij}(0) .
\label{19}
\end{equation}

\noindent Global phases do not contribute to the density matrix. Eq. (\ref{19})
is compatible with eq. (\ref{17}) since the element  $\rho^H_{ij}(t)$
is only different from zero if in the
initial state we already have a superposition of the $i^{th}$ and $j^{th}$
instantaneous eigenstates of ${H}(t)$. Finally, the difference of phases, either the
geometrical and the dynamical phases, appears naturally in this formalism.

We notice that eq. (\ref{19}) includes the  elements $\rho^H_{iN}(t)$,
$i= 1,2,\cdots, N-1$, as well as the solution (\ref{17}) for the
diagonal terms. The inhomogeneous term in eq. (\ref{10}) gives
a contribution of order ($\frac{1}{T}$); this is why it does not
contribute to the dynamics of the  elements $\tilde{\rho}_{iN}(s)$
in the adiabatic limit.

The argument of the first exponential on the r.h.s. of eq.
(\ref{19}) is equal to the difference of the geometric phases
acquired by the instantaneous eigenstates $i$ and $j$ of the
Hamiltonian $ H(t)$. For any closed path in the $\vec{\bf R}$-space,
each of those phases depends only on the path followed by 
the $\vec{\bf R}$-parameters. As should be, in the density matrix formalism
 we recover the known  results derived for the 
unitary evolution  of instantaneous eigenstates  of Hamiltonian
 in the adiabatic limit\cite{berry}.


\section{The Quantum System in the Presence of Weak \\
        Dissipation}

In general we are interested in studying a quantum system that is
part of a whole system whose sub-systems interact with one another.
This interaction allows the sub-systems to have exchanges
among themselves. The traditional way to study a part
of the whole system is taking a partial trace over
all degrees of freedom of the complementary sub-system. These
complementary degrees of freedom are called environment. In this approach
we have an effective Hamiltonian that drives the
dynamical evolution of the quantum system under study and at the same time
the non-unitary part of the Liouvillian takes into
account its interaction with the  environment. In the general case
the effective Hamiltonian depends on a set of time-dependent
classical parameters.

In reference \cite{JphysA} we questioned whether the
presence of dissipation could introduce an imaginary geometric
phase in systems driven by periodic Hamiltonians. We considered a
two-level model in the presence of two distinct  Lindblad operators
representing reservoirs, and concluded that in those
 cases the imaginary phases are {\bf not} geometric and, based on
eq. (19) of reference \cite{JphysA}, we affirmed that the nature
of the imaginary phase depends on its origin. Certainly, this last affirmation
has to be confirmed by a model-independent approach.

The last term on the r.h.s. of eq. (\ref{04}) introduces the effects of dissipation in
the dynamics of the quantum system. We assume that the dissipation is
weakly coupled to the quantum system. In general, the non-unitary part
of the Liouvillian is written in a time-independent basis. Let $\{ | v_l
\rangle \}$ be this time-independent basis and, for weakly dissipative
interaction, we have \cite{13,14}

\begin{equation}
\langle v_i | {\cal L}_D \rho (t) | v_j \rangle =
\sum_{l,m} \; {\bar c}^{ij}_{ml} (t) \rho_{lm}(t),
\end{equation}

\noindent where $\rho_{lm}(t) \equiv \langle v_l | \rho (t)
| v_m \rangle$. The coefficients ${\bar c}^{ij}_{ml} (t)$
take into account the characteristics of the environment
(in the case of a quantum system at thermodynamic
equilibrium with a  reservoir composed of an infinity 
set of harmonic oscillators, the coefficients ${\bar c}^{ij}_{ml} (t)$
take into account the distribution of frequencies, etc, but
are time-independent). In the most
general case of coupling with a dissipative medium we can be
time dependent and consequently  those coefficients
can vary in time.

From the beginning we chose to write the density matrix in the
basis of the instantaneous eigenstates of
Hamiltonian $ H(t)$. In this basis, the non-unitary part of Liouvillian
is written as

\begin{equation}
\langle u_i;t | {\cal L}_D \rho (t) | u_j;t\rangle =
\sum_{l,m} \; c^{ij}_{ml} (t) \rho^H_{lm}(t).
\label{40}
\end{equation}

\noindent The coefficients $c^{ij}_{lm} (t)$ on the r.h.s. of eq.
(\ref{40}) are obtained from  ${\bar c}^{ij}_{ml} (t)$ by
making a similarity transformation in each of its indices, that is

\begin{equation}
c^{ij}_{lm} (t) = \langle u_i;t | v_{l_1} \rangle \langle u_l;t | v_{l_3} \rangle
{\bar c}^{l_1 l_2}_{l_3 l_4} (t) \langle v_{l_2} | u_j;t \rangle
\langle v_{l_4} | u_m;t \rangle ,
\label{44.03}
\end{equation}

\noindent where implicit sum over the indices $l_1$, $l_2$, $l_3$ and
$l_4$ is meant. We point out that even in the case when the coefficients 
${\bar c}^{l_1 l_2}_{l_3 l_4}$ are time independent, their analogous in the
instantaneous basis can acquire a time dependence  through the transformation
(\ref{44.03}). From  the scalar products in the transformation
(\ref{44.03}) the coefficients $c^{ij}_{lm} (t)$ do not 
get a dependence  on the variation
of Hamiltonian or of any other external classical parameter.

Following the same steps as we did in section 2,
we obtain the dynamical equations for the
elements $\tilde{\rho}_{ij}(s)$ (see their definition in eq.
(\ref{2})) in the limit $T \rightarrow \infty$:

\vspace{.3cm}

\noindent {\it i)} for the diagonal elements $\tilde{\rho}_{ii}(s)$,
$i=1,2,\cdots, N-1$.

\begin{eqnarray}
\frac{d\tilde{\rho}_{ii} (t)}{dt} = \sum_{l=1}^{N-1} \Big( c^{ii}_{ll}(k(t))
- c^{ii}_{NN}(k(t))\Big) \tilde{\rho}_{ll} (t) +
c^{ii}_{NN} (k(t)) .
\label{41}
\end{eqnarray}

\noindent In writing the r.h.s. of eq. (\ref{41}) we have already implemented  the
condition Tr$(\rho (t)) =1$. In the most general case the
dissipation couples the dynamics of the diagonal elements of the density
matrix.  From eq. (\ref{41}) we recover the adiabatic behavior
of the quantum system in the presence of the  dissipative medium
when the non-unitary part of the Liouvillian  has
null  diagonal terms in the instantaneous basis ($c_{ll}^{ii} = 0$).
In this situation the population of an instantaneous eigenstate of Hamiltonian
does not vary along the adiabatic process. In this
case the quantum system does not transfer to the environment energy due to
electronic transitions.

The elements of matrix ${\bf C}(t)$  are defined as

\begin{equation}
C_{il}(t) \equiv c^{ii}_{ll}(t) -  c^{ii}_{NN} (t),
\label{41.1}
\end{equation}

\noindent $i$, $l = 1,2, \cdots , N-1$. The general solution of
eq. (\ref{41}) is

\begin{equation}
\rho^H_{ii}(t) = \Big[{\cal T}\Big( e^{\int_0^t dt^\prime
{\bf C}(t^\prime)}\Big) \Big]_{ij} \Big\{ \int_0^t dt^\prime
\Big[ {\cal T}\Big( e^{\int_{t^\prime}^0 dt^{\prime\prime}
{\bf C}(t^{\prime\prime})}\Big) \Big]_{jk} c^{kk}_{NN} (t^\prime)
+ \rho^H_{jj}(0) \Big\} .
\label{41.2}
\end{equation}

\noindent  The operator
${\cal T}$ means a time-ordering operator \cite{time_operator}. From eq. (\ref{44.03})
we obtain that the coefficients $c^{ii}_{ll}(t)$, $i,l=1,2,\cdots,N$, are the 
same for any basis of the instantaneous eigenstates of Hamiltonian. This fact
avoids any ambiguity in the imaginary phases in the time-ordering terms.
In the general case the imaginary phases in eq. (\ref{41.2}) are time-dependent.
We postpone the discussion under what conditions the time-ordering integrals in 
eq. (\ref{41.2}) can be rewritten as a path integrals in a suitable parameter
space.

From eq. (\ref{41.2}) we recover solution (\ref{17}) in the
absence of dissipation ($c^{ii}_{ll} =0$, $i,l= 1,..., N$).

\vspace{.3cm}

\noindent {\it ii)} for the elements $\tilde{\rho}_{ij}(s)$,
$i=1,2,\cdots,N-1$ and $j=i+1,\cdots,N$.

\begin{eqnarray}
\frac{d\tilde{\rho}_{ij} (t)}{dt} &=&
\Big[ - \langle u_i ; t| \frac{d | u_i ; t \rangle}{dt} +
\langle u_j ; t| \frac{d | u_j ; t \rangle}{dt} +
c^{ij}_{ij}(k(t)) \Big]
\tilde{\rho}_{ij} (t) + \nonumber\\
\nonumber \\
&+& \sum_{\{l,m\} \atop{l \neq i, m \neq j}} c^{ij}_{ml} (k(t)) \tilde{\rho}_{lm}(t),
\label{43}
\end{eqnarray}

\noindent where the set of pair of indices  $\{l,m\}$ are those which satisfy:
$E_{l}(t) - E_{m}(t)= E_i(t) - E_j(t)$. To simplify our
discussion, we order the instantaneous eigenenergies such that: if
$E_i(t) < E_j(t)$ then $i<j$. Since the indices ($i$, $j$) of the
elements of the density matrix in eq. (\ref{43}) are chosen such that
$i<j$, we get that the elements $\tilde{\rho}_{lm}(t)$ that contribute
to the r.h.s. of this equation are such that $l<m$. Once
the energy spectrum of the quantum system is non-degenerate, if the
elements $\tilde{\rho}_{l_1 m_1}(t)$ and $\tilde{\rho}_{l_2 m_2}(t)$
that contribute to r.h.s. of eq. (\ref{43}) are distinct  then  we must necessarily
have $l_1 \neq l_2$ and $m_1 \neq m_2$.

Let us suppose that the dynamics of $M$ elements $\tilde{\rho}_{ij}(t)$
are coupled by the presence of dissipation and their dynamics 
are given by eq. (\ref{43}).
Due to the fact that each pair ($l, m$) is unique we may
relabel them by using only one index: ($l_i, m_i$), $i=1$, $2$, $\cdots$, $M$.
Eq. (\ref{43}) is rewritten as:

\begin{equation}
\frac{d \tilde{\rho}_{l_i m_i}(t)}{dt} = \Big[ - \langle u_{l_i};t| \Big(
\frac{d}{dt} | u_{l_i};t \rangle \Big) + \langle u_{m_i};t| \Big(
\frac{d}{dt} | u_{m_i};t \rangle \Big) \Big] \tilde{\rho}_{l_i m_i}(t) +
\sum_{k=1}^M c^{l_i m_i}_{m_k l_k} \tilde{\rho}_{l_k m_k}(t),
     \label{43.1}
\end{equation}

\noindent $i= 1, 2,..., M$. We distinguish two possible situations:

\vspace{.2cm}

\noindent {\it 1)} the dynamics of the elements $\tilde{\rho}_{ij}(t)$ are not
coupled by the coupling to a weakly dissipative medium: $M=1$.

In this case, eq. (\ref{43.1}) reduces to

\begin{equation}
\frac{d \tilde{\rho}_{ij}(t)}{dt} = \Big[ - \langle u_i;t| \Big(
\frac{d}{dt} | u_i;t \rangle \Big) + \langle u_j;t| \Big(
\frac{d}{dt} | u_j;t \rangle \Big) \Big] \tilde{\rho}_{ij}(t) +
c^{ij}_{ji} \tilde{\rho}_{ij}(t),
   \label{43.2}
\end{equation}

\noindent whose solution is (see eq. (\ref{2}))

\begin{eqnarray}
\rho^H_{ij} (t) =
e^{-i\int_0^t dt^\prime (E_i(t^\prime)- E_j(t^\prime))}
e^{- \int_0^t dt^\prime \Big[ \langle u_i ; t^\prime| \frac{d | u_i ;
t^\prime \rangle}{dt^\prime} -
\langle u_j ; t^\prime| \frac{d | u_j ; t^\prime \rangle}{dt^\prime} \Big]}
e^{\int_0^t dt^\prime c^{ij}_{ji}(t^\prime)}\rho^H_{ij} (0) .
\label{43.3}
\end{eqnarray}

\noindent Here again the coefficients $c^{ij}_{ji}(t)$ that appear in the last 
phase on the r.h.s. of eq. (\ref{43.3}) are independent of the chosen basis of the instantaneous
eigenstates of Hamiltonian. In the general case this phase is time-dependent.
In next sub-section we discuss the conditions satisfied by the coefficients $c^{ij}_{ji}(t)$ 
in order to this integral becomes path-dependent.

Eq. (\ref{43.3}) reduces to eq. (\ref{19}) in the absence of
interaction with a dissipative medium.

\vspace{.2cm}

\noindent {\it 2)} the dynamics of $M$ elements $\tilde{\rho}_{ij}(t)$ are coupled due
to the presence of dissipation: $M>1$.

Making the change of variables

\begin{equation}
\bar{\rho}_{l_i m_i}(t) \equiv
e^{ \int_0^t dt^\prime \Big[ \langle u_{l_i} ;
t^\prime| \frac{d | u_{l_i} ; t^\prime \rangle}{dt^\prime} -
\langle u_{m_i} ; t^\prime| \frac{d | u_{m_i} ; t^\prime \rangle}{dt^\prime} \Big]}
\tilde{\rho}_{l_i m_i}(t),  \label{43.31}
\end{equation}

\noindent where $i=1, 2, \cdots , M$, eq. (\ref{43.1}) becomes

\begin{equation}
\frac{d \bar{\rho}_{l_i m_i}(t)}{dt} = A_{ik}(t) \bar{\rho}_{l_k m_k}(t).
\label{43.4}
\end{equation}

\noindent The elements $A_{ik}(t)$ of  matrix ${\bf A}(t)$ are defined as

\begin{equation}
A_{ik}(t) \equiv c^{l_i m_i}_{m_k l_k} (t)
e^{- \int_0^t dt^\prime \Big[ \langle u_{l_k} ;
t^\prime| \frac{d | u_{l_k} ; t^\prime \rangle}{dt^\prime} -
\langle u_{l_i} ; t^\prime| \frac{d | u_{l_i} ; t^\prime \rangle}{dt^\prime} -  \langle u_{m_k} ;
t^\prime| \frac{d | u_{m_k} ; t^\prime \rangle}{dt^\prime} +
\langle u_{m_i} ; t^\prime| \frac{d | u_{m_i} ; t^\prime \rangle}{dt^\prime} \Big]}.
\end{equation}

The solution of eq. (\ref{43.4}) is

\begin{equation}
\bar{\rho}_{l_i m_i}(t) = \Big[ {\cal T} \Big( e^{\int_0^t {\bf A} (t^\prime)
dt^\prime} \Big) \Big]_{ij} \bar{\rho}_{l_j m_j}(0),
\end{equation}

\noindent ${\cal T}$ being the time-ordering operator \cite{time_operator}.

From eqs. (\ref{2}) and (\ref{43.31}) we finally have

\begin{eqnarray}
\rho^H_{l_i m_i} (t) &=&
e^{-i\int_0^t dt^\prime (E_{l_i}(t^\prime)- E_{m_i}(t^\prime))}
e^{- \int_0^t dt^\prime \Big[ \langle u_{l_i} ; t^\prime| \frac{d | u_{l_i} ;
t^\prime \rangle}{dt^\prime} -
\langle u_{m_i} ; t^\prime| \frac{d | u_{m_i} ; t^\prime \rangle}{dt^\prime} \Big]}
\times \nonumber \\
\nonumber \\
&\times& \Big[ {\cal T} \Big( e^{\int_0^t {\bf A} (t^\prime)
dt^\prime} \Big) \Big]_{ij} {\rho}^H_{l_j m_j}(0).
\label{43.5}
\end{eqnarray}

\noindent  By choosing a new basis of eigenstates of Hamiltonian the 
elements $A_{ik}(t)$ get an irrelevant real phase that does not contribute 
to the average value of any physical operator. As in the two previous 
discussion, the time-ordering integral that appears on the r.h.s. of 
eq. (\ref{43.5}) due to the presence of dissipation is time-dependent. In the next
sub-section we give the conditions necessary for this imaginary phase to be 
path-dependent of a suitable set of time dependent parameters.

Eq. (\ref{19}) is recovered from eq. (\ref{43.5})
in the absence of a coupling with a dissipative medium.


\subsection{Conditions to Obtain an Imaginary Geometric Phase}

In the most general case, eqs. (\ref{41.2}), (\ref{43.3}) and (\ref{43.5})
do not give an imaginary geometric phase correction
 to the real Berry phase\cite{berry}.
In this subsection we discuss the mathematical requirements  that the
coefficients $c^{ij}_{mk}(k(t))$ have to satisfy in order to those
imaginary phases are geometric (path-dependent).

Differently from the real phases acquired by the evolution of the instantaneous 
eigenstates of Hamiltonian in the adiabatic limit\cite{born, adia_theor, adia_theor1}, 
the imaginary phases in eqs. (\ref{41.2}), (\ref{43.3}) and (\ref{43.5}) have no
ambiguity due to the arbitrariness of the basis of the instantaneous eigenstates 
of Hamiltonian. Consequently in the case of the imaginary phases acquired by the
elements of the density matrix $\rho^H(t)$ due to the presence
of dissipation, they can be written as an integral
over a suitable time-dependent parameter space if we have an integration over $t$
of a function $f(t)$ that has the general form

\begin{equation}
f(t) = \sum_i \varphi_i(t) \frac{d \Psi_i(t)}{dt},
\label{44.1}
\end{equation}

\noindent with the functions $\Psi_i(t)$ satisfying the
following conditions:

\vspace{.3cm}

\noindent {\bf 1)} the functions $\Psi_i(t)$ are not
explicitly time-dependent;

\vspace{.2cm}

\noindent {\bf 2)} the time-dependence of functions
$\Psi_i(t)$ come only from their dependence on the set
of parameters $\vec{{\bf k}} (t)\equiv ( k_1(t),k_2(t), \cdots, k_l(t))$.

\vspace{.3cm}

We point out that we do no restrict the regime of the time variation of the set of
parameters $\vec{\bf k}(t)$ and it has not to be a periodic function in time.

We begin by discussing the imaginary phase on
 the r.h.s. of eq. (\ref{41.2}). On the
 r.h.s. of eq. (\ref{41.2}) we have a time-ordering integrals of
 a matrix ${\bf C}(t)$ whose elements are (see eq. (\ref{41.1})):

\begin{equation}
C_{il}(t) = c^{ii}_{ll}(t) -  c^{ii}_{NN} (t),  \hspace{2cm}
i, l = 1,2, \cdots , N-1.
    \label{44.2}
\end{equation}

\noindent  The relation between $c^{ii}_{ll}(t)$ and the
coefficients of the non-unitary part of Liouvillian written
in a time-independent basis is given by eq. (\ref{44.03})

\begin{equation}
c^{ii}_{ll} (t) = \langle u_i;t | v_{l_1} \rangle \langle u_l;t | v_{l_3} \rangle
{\bar c}^{l_1 l_2}_{l_3 l_4} (t) \langle v_{l_2} | u_i;t \rangle
\langle v_{l_4} | u_l;t \rangle,
            \label{44.3}
\end{equation}

\noindent where $i$, $l= 1,2,\cdots, N$. In order to be able
to write the time-ordering integrals on the r.h.s. of eq.
(\ref{41.2}) as a path-dependent integral
the coefficients ${\bar c}^{l_1 l_2}_{l_3 l_4} (t) $
must have the form

\begin{equation}
{\bar c}^{l_1 l_2}_{l_3 l_4} (t) = \frac{d}{dt}\Big(\Psi^{l_1 l_2}_{l_3 l_4}
(t)\Big),
\label{44.4}
\end{equation}

\noindent and the functions $\Psi^{l_1 l_2}_{l_3 l_4}(t)$ have to
satisfy the two previous conditions mentioned.  We stress out that the
path in the $\vec{\bf k}$-parameter space has not to be closed.

Due to the hermicity property of the operator ${\cal L}_D \rho(t)$ the 
elements of matrix ${\bf C}$ (see eq.(\ref{41.1})) are real. If condition 
(\ref{44.4}) is satisfied the geometric phase introduced by the presence 
of dissipation is purely imaginary.

\vspace{.5cm}

Once eq. (\ref{44.4}) is valid for the coefficients
${\bar c}^{l_1 l_2}_{l_3 l_4} (t)$, the non-diagonal elements
$\rho^H_{lm}(t)$ have two distinct situations in the limit $T \rightarrow \infty$:

\vspace{.3cm}

\noindent {\it i)} for $M=1$, the elements $\rho^H_{lm}(t)$ get an
imaginary geometric phase. From the hermicity property of operator ${\cal L}_D
\rho (t)$ we obtain that $c^{ij}_{ji}(t) = \Big(c^{ji}_{ij}(t)\Big)^*$ which
leaves open the possibility that the coefficient $c^{ij}_{ji}(t)$ has an imaginary
part. For master equations where $c^{ij}_{ji}(t)$ has an imaginary part, the
dissipation gives a real correction to the Berry's phase\cite{berry}. This correction
to the Berry's phase has not to be a closed-loop in the $\vec{\bf k}$-parameter space.

\vspace{.2cm}

\noindent {\it ii)} for $M>1$, the elements $\rho^H_{lm}(t)$ get
an imaginary geometric phase due to the interaction with
a weakly dissipative medium. The presence of the dissipation couples the dynamics
of different elements of the density matrix and the integral over the
$\vec{\bf k}$-path involves a matrix that in general does not commute with itself at
different instants. This $\vec{\bf k}$-path has not to be a closed loop. As in
case {\it i} if the coefficients $c^{ij}_{lm}(t)$ have an imaginary part they
give a correction to the real geometric phase\cite{berry}.

From eq. (\ref{44.4}) we see that the presence of the dissipative media
gives an imaginary geometric phase only if the coupling between the quantum
system and its environment is externally driven by a varying external 
parameter.  In both situations the time-variation of the set
of the parameters $\vec{\bf k}(t)$ has not to be slow.


\section{Adiabatic Limit  of the Density Matriz of \hfill \\
         the Spin $\frac{1}{2}$ Model Coupled to a Reservoir }

For systems in contact with reservoirs at thermal equilibrium
whose  the weak coupling constant does not vary in time,
the  effect of the  presence of the dissipation  is to destroy
the coherence in the quantum system in a time-dependent 
process. That is the case of two distinct reservoirs coupled
to a spin $\frac{1}{2}$ model studied in detail in reference \cite{JphysA}.

By using directly eq. (\ref{44.4}) we want to verify
in the next sub-sections  if the 
conclusions about the nature of the imaginary phases
in reference \cite{JphysA} are corrected or not.


\subsection{Dephasing Process in  a Two-Level System}

An interesting process well studied in the standard
textbooks\cite{14,lindblad} is the phase destroying
process which might appear due to elastic collisions.
We consider a spin $\frac{1}{2}$ variable (two level model) coupled to a time dependent
magnetic field  precessing around the z-axis with $\omega$ constant
precession frequency. The external magnetic field $\vec{B}(t)$ has norm $B$ and
makes a $\theta$ angle with the $z$-axis.

For the sake
of later calculations, it is convenient to define two unitary
transformations: the first one, $R(\omega, t)$, takes us to the rotating
frame where the Hamiltonian is no  longer time dependent;
the second one, $D(B, \theta,\omega)$, diagonalizes
the effective Hamiltonian (time independent) that drives the dynamics of the
final matrix representation of the density operator.
We call this the diagonal frame \cite{JphysA}.

The master equation to this process written in the diagonal frame is

\begin{equation}
\frac{d}{dt}\rho_D\left( t\right) =-{i}
\left[ \lambda_1 {\sigma }_{z},\rho_D\left( t\right) \right]
+ \frac{k}{2}  \left( \sigma_z \rho_D (t) \sigma_z  - \rho_D (t) \right),
                      \label{ap_4}
\end{equation}

\noindent where $k$ is the dissipation constant at zero temperature.
The weak coupling regime
 is characterized by the condition $ \frac{k}{\lambda_1} \ll 1$
with $\lambda_1 = \sqrt{ (\mu B \cos(\theta)- \frac{\omega}{2})^2
       + \mu^2 B^2 \sin^2 (\theta)}$.

We define $\rho^H(t)$ to be  the density matrix in a basis of the
instantaneous eigenvectors of Hamiltonian.
The relation between  $\rho^H(t)$ and $\rho_D(t)$ is \cite{JphysA}

\begin{subequations}\label{ap_40}

\begin{equation}
\rho^H(t) = {\bf V}^\dag(t)\,{\bf D}\,  \rho_D(t) \, {\bf D} \, {\bf V}(t),
           \label{ap_40_1}
\end{equation}

\noindent where the matrix ${\bf V}(t)$ is equal to

\begin{equation}
 {\bf V}(t) = \left( \begin{array}{lr}
                    \cos(\frac{\theta}{2}) e^{-\frac{i\omega t}{2}} &
                                         - \sin(\frac{\theta}{2}) e^{-\frac{i\omega t}{2}} \\
                     \sin(\frac{\theta}{2}) e^{\frac{i\omega t}{2}} &
                     \cos(\frac{\theta}{2}) e^{\frac{i\omega t}{2}}
                \end{array}\right)    \label{ap_40_2}
\end{equation}

\noindent and

\begin {equation}
{\bf D} = \frac{1}{\sqrt{2}} \sqrt{1+\Lambda} \; \sigma_z 
      + \frac{1}{\sqrt{2}} \sqrt{1-\Lambda} \; \sigma_x    \label{ap_40_3}
\end{equation}

\end{subequations}

\noindent with $\Lambda \equiv \frac{1}{\lambda_1}
           \Big(\mu B \cos(\theta) - \frac{\omega}{2}\Big)$
and $\pm \lambda_1$ are the eigenvalues of the effective Hamiltonian.

\vspace{0.2cm}

The master equation (\ref{ap_4}) in the instantaneous basis of the
Hamiltonian for arbitrary value of $\omega$ is

\begin{subequations}\label{ap_5}

\begin{equation}
\frac{d}{dt}\rho^H\left( t\right) = -{i} \left[
\left( \mu B + \frac{\omega}{2} \right) {\sigma }_{z} -  \frac{\omega}{2} \sigma_n(t) ,
                \rho^H\left( t\right) \right]
                           +\frac{k}{2} {\cal L}_D^H \rho^H (t)\label{ap_5_1}
\end{equation}

\noindent where
\vspace{-0.6cm}

\begin{eqnarray}
{\cal L}_D^H \rho^H (t) &=& \Lambda^2 \sigma_n(t) \rho^H (t) \sigma_n(t) +
                          \Lambda\sqrt{1-\Lambda^2} \Big[
                          e^{-i\omega t} \Big( \sigma_n(t) \rho^H (t) \sigma_+(t) +
                          \sigma_+(t) \rho^H (t) \sigma_n(t) \Big) +
                                 \nonumber \\
                    &+& e^{i\omega t} \Big( \sigma_n(t) \rho^H (t) \sigma_-(t) +
                        \sigma_-(t) \rho^H (t) \sigma_n(t) \Big) \Big] +
                        (1-\Lambda^2)\Big[
                        e^{-2i\omega t} \sigma_+(t) \rho^H (t) \sigma_+(t) +
                              \nonumber\\
                    &+& e^{2i\omega t} \sigma_-(t) \rho^H (t) \sigma_-(t) +
                        \sigma_+(t) \rho^H (t) \sigma_-(t) +
                        \sigma_-(t) \rho^H (t) \sigma_+(t) \Big] - \rho^H(t).
                    \label{ap_5_2}
\end{eqnarray}

\end{subequations}

\noindent In  eqs. (\ref{ap_5}), we use the definitions:

\begin{subequations}

\begin{equation}
 \sigma_n(t) \equiv \left( \begin{array}{lr}
                    \cos(\theta) &  - \sin(\theta) e^{-i\omega t} \\
                     - \sin(\theta) e^{i\omega t}   & - \cos(\theta)
                \end{array}\right) ,\label{ap_5_3}
\end{equation}

\begin{equation}
\sigma_+(t) \equiv e^{i\omega t}
\left( \begin{array}{lr}
     \frac{\sin(\theta)}{2}&
     \cos^2(\frac{\theta}{2}) e^{-i\omega t} \\
       -\sin^2(\frac{\theta}{2}) e^{i\omega t}
          & - \frac{\sin(\theta)}{2}
  \end{array}\right) ,      \label{ap_5_4}
\end{equation}

\noindent and  $\sigma_-(t) \equiv \Big(\sigma_+(t)\Big)^\dagger$.

\end{subequations}

From eqs. (\ref{41.2}) and (\ref{43.3}) we see that to obtain the nature of the imaginary
phase (if they are time or path-dependent) we need the coefficients $c_{lm}^{11}$
and $c_{lm}^{12}$ gotten from ${\cal L}_D^H \rho^H (t)$ in the adiabatic 
limit ($\omega \rightarrow 0$).  However, the weakness condition on the
dissipation constant $k$ ($\frac{k}{\lambda_1}$) does not impose any constraint
to the ratio $\frac{k}{\omega}$.  The interesting case happens when the adiabatic 
evolution and the dissipation effect are of the same order: $\frac{k}{\omega} \sim 1$.

In the adiabatic limit, weak dissipation regime and for
$\frac{k}{\omega} \sim 1$, eq. (\ref{ap_5_2}) give us

\vspace{-.4cm}

\begin{subequations}\label{dom_44}
\begin{eqnarray}
c^{11}_{11}=k \;{\cal O}(\frac{\omega}{\mu B})   \hspace{1cm}
{\rm and} \hspace{1cm}
c^{11}_{22}=k \;{\cal O}(\frac{\omega}{\mu B}) \label{dom_44a}
\end{eqnarray} 

\vspace{-.4cm}
 
 \noindent and
 
 \vspace{-.4cm}
 
\begin{eqnarray}
c^{12}_{21}= - k\Big[1 +{\cal O}(\frac{\omega}{\mu B})\Big].
\label{dom_44b}   
\end{eqnarray}
\end{subequations}

\noindent We neglect the contribution of terms of order 
$k \hspace{.06cm} {\cal O}(\frac{\omega}{\mu B})$ to eqs. (\ref{41.2}) and
(\ref{43.3}) since they are of the same magnitude as the terms of higher order
in $(\frac{1}{T})$. The constant $k$  can not be 
written as eq. (\ref{44.4}) and consequently  the imaginary phase due
 to the weak coupling to the dissipative medium is {\bf not} geometric. 
For completeness we substitute the values of the coefficients in eqs. (\ref{41.2})
and (\ref{43.3}), and obtain\cite{nota}

\vspace{-.4cm}

\begin{subequations}\label{dom_45}
\begin{equation}
\rho^H_{11}(t)=\rho^H_{11}(0) \label{dom_45a}
\end{equation}

\vspace{-.4cm}

\begin{equation}
\rho^H_{12}(t)= e^{-2 i \mu B t} e^{-i\omega(1-\cos(\theta))t}
e^{-k t}\rho^H_{12}(0).
\label{dom_45b}
\end{equation}

\end{subequations}

\noindent From eq. (\ref{dom_45a}) we get that the process continues to be
adiabatic even in the presence of dissipation. On the r.h.s. of eq. (\ref{dom_45b})
the first phase gives the difference of the dynamical phases of the eigenstates of 
Hamiltonian, the second phase gives the difference of the geometrical phases 
acquired by the instantaneous eigenstates of ${H}(t)$ and the last phase 
is the time-dependent phase due to the coupling of the quantum system
with the dissipative medium. Its
effect is to destroy  the quantum coherence  in the system.


\subsection{ Adiabatic Limit of a Two-Level Model in Thermal \\
                         Equilibrium }

In our next application, we consider the spin $\frac{1}{2}$  model under the influence
 of an external magnetic field as 
described in section 4.1 but now coupled to a reservoir of electromagnetic fields
at thermal equilibrium \cite{JphysA}. The master equation of this model  in
thermodynamic equilibrium in the diagonal frame is \cite{14,lindblad}

\begin{eqnarray}
\frac{d}{dt}\rho_D\left( t\right) &=&-{i} \left[
\lambda_1 {\sigma }_{z},\rho_D\left( t\right) \right]
+k\left( \overline{n}+1\right) \left( 2{\sigma }_{-}\rho_D
\left( t\right) {\sigma }_{+}-\rho_D\left( t\right)
{\sigma }_{+}{\sigma }_{-}-{\sigma }_{+}
{\sigma }_{-}\rho_D\left( t\right) \right) + \nonumber \\
%
%
&& + k\overline{n}\left( 2{\sigma }_{+}\rho_D\left( t\right)
{\sigma }_{-}-\rho_D\left( t\right) {\sigma }_{-}
{\sigma }_{+}-{\sigma }_{-}{\sigma }_{+}
\rho_D \left( t\right) \right),
           \label{ap_10}
\end{eqnarray}

\noindent where $k$ is the dissipation constant at zero temperature
and $\bar{n}$ is the average number of excitations of the weakly
coupled thermal oscillators at inverse temperature $\beta$.

By doing the transformation (\ref{ap_40}),
the master equation (\ref{ap_10}) in the instantaneous basis of the
Hamiltonian for arbitrary value of $\omega$ becomes

\begin{equation}
\frac{d}{dt}\rho^H\left( t\right) = -{i} \left[
\left( \mu B + \frac{\omega}{2} \right) {\sigma }_{z} -  \frac{\omega}{2} \sigma_n(t) ,
                \rho^H\left( t\right) \right]
                           +k {\cal L}_D^H \rho^H (t)\label{A1.1}
\end{equation}

\noindent where
\vspace{-0.6cm}

\begin{eqnarray}
{\cal L}_D^H \rho^H (t) &=& \frac{2\overline{n}+1}{2}\Big\{ -2 \rho^H (t) +
                          (1 - \Lambda^2)\Big[ \sigma_n(t) \rho^H (t) \sigma_n(t) -
                          e^{-2i\omega t} \sigma_+ (t) \rho^H (t) \sigma_+(t) -
                         \nonumber \\
             &-&  e^{2i\omega t} \sigma_- (t) \rho^H (t) \sigma_-(t) \Big]
             +(1+\Lambda^2)\Big[\sigma_+ (t) \rho^H (t) \sigma_-(t) +
                  \sigma_- (t) \rho^H (t) \sigma_+(t)\Big] -
             \nonumber \\
             &-& \Lambda\sqrt{1-\Lambda^2}
              \Big[e^{i\omega t} \Big( \sigma_n(t) \rho^H (t) \sigma_-(t)
               + \sigma_- (t) \rho^H (t) \sigma_n(t) \Big) +
              \nonumber\\
             &+& e^{-i\omega t} \Big( \sigma_+ (t) \rho^H (t) \sigma_n(t) +
                 \sigma_n (t) \rho^H (t) \sigma_+(t) \Big)\Big]\Big\} \; -
                 \nonumber \\
             &-& \frac{1}{2} \Big\{ \{\rho^H (t), \Lambda \sigma_n(t) +
                    \sqrt{1-\Lambda^2}(e^{-i\omega t}\sigma_+(t) +
                    e^{i\omega t}\sigma_-(t)\} + 2\Lambda\Big[
                    \sigma_+ (t) \rho^H (t) \sigma_-(t) -
                    \nonumber\\
             &-& \sigma_- (t) \rho^H (t) \sigma_+(t)\Big] + \sqrt{1-\Lambda^2}
             \Big[ e^{-i\omega t} \Big( \sigma_n(t) \rho^H (t) \sigma_+(t) -
             \sigma_+(t) \rho^H (t) \sigma_n(t) \Big) +
               \nonumber\\
            &+& e^{i\omega t} \Big( \sigma_-(t) \rho^H (t) \sigma_n(t) -
            \sigma_n(t) \rho^H (t) \sigma_-(t) \Big)\Big]\Big\} ,
            \label{A1.2}
\end{eqnarray}

\noindent $\sigma_n(t)$ and $\sigma_+(t)$ are given by eqs. (\ref{ap_5_3}) and (\ref{ap_5_4})
respectively.

As in sub-section 4.1, we consider the adiabatic limit and the weak dissipative regime.  
In those regimes we consider the case  $\frac{k}{\omega} \sim 1$
and the coefficients  $c_{lm}^{11}$ and $c_{lm}^{12}$ gotten 
from ${\cal L}_D^H \rho^H (t)$ (see eq. (\ref{A1.2})) are:

\begin{subequations} \label{dom_48}

\begin{equation}
c^{11}_{11} = k \Big[ -2(1+\overline{n}) + {\cal O} (\frac{\omega}{\mu B})\Big]
\hspace{1cm} {\rm and} \hspace{1cm} 
c^{11}_{22}=  k \Big[ 2 \overline{n} + {\cal O} (\frac{\omega}{\mu B})\Big]
\label{dom_48a}
\end{equation}

\noindent and

\vspace{-.4cm}

\begin{eqnarray}
c^{12}_{21}=  k \Big[ -(1+2 \overline{n}) + {\cal O} (\frac{\omega}{\mu B})\Big].
\label{dom_48b}
\end{eqnarray}

\end{subequations}

\noindent By the reason discussed in sub-section 4.1 the terms $k \hspace{.6cm}
{\cal O} (\frac{\omega}{\mu B})$ do not contribute to the adiabatic 
limit. For the present model we see that neither the  coefficients
$c^{11}_{lm}$ and $c^{12}_{lm}$ can be written as eq. (\ref{44.4})
and consequently the phases due to the coupling  of the spin $\frac{1}{2}$ model  with
a reservoir of electromagnetic fields at thermal equilibrium are {\bf not} 
geometric.

The solutions of eqs. (\ref{41.2}) and (\ref{43.3}) for this model
are \cite{nota}

\begin{subequations} \label{dom_49}

\begin{equation}
\rho^H_{11}(t)= \Big[ \rho^H_{11}(0) - \frac{\overline{n}}{1+ 
2 \overline{n}}\Big] e^{-2k(1+2 \overline{n})t} + 
\frac{\overline{n}}{1+2 \overline{n}}
\label{dom_49a}
\end{equation}

\noindent and 

\begin{equation}
\rho^H_{12}(t)=e^{-2 i \mu B t} e^{-i \omega(1-\cos(\theta))t}
e^{-k(1+2 \overline{n})t}\rho^H_{12}(0).
\label{dom_49b}
\end{equation}

\end{subequations}

\noindent From eq. (\ref{dom_49a}) we get that due to the weak interaction
with the dissipative medium, we have a very slow phenomena but the adiabatic 
character of the quantum process is lost. As in the previous sub-section,
the constants $k$ and ${\overline{n}}$ can not be 
written as eq. (\ref{44.4}) and consequently  the imaginary phase due
 to the weak coupling to the dissipative medium is {\bf not} geometric. 
From eq. (\ref{dom_49b}) we obtain that the interference
effects between the instantaneous eigenstates of Hamiltonian is destroyed
by the coupling with the dissipative medium. This effect depends on the 
dissipative constant $k$ and the time elapsed.


\section{Conclusions}

We discuss the behavior of  the density matrix
of  non-degenerate quantum systems whose dynamics are driven 
by a master equation in which the unitary part is a periodic
Hamiltonian with period $T$ ($T \rightarrow \infty$). Our discussion
is model-independent. We recover the known results of the literature
\cite{berry} for the quantum systems that are not coupled to
any external dissipative medium. The  difference of geometric phases 
 necessary for measuring a physical effect associated
to the existence of these phases appears naturally 
in the density matrix approach. The obtainment of the adiabatic
limit for the density matrix in a closed system is used as  a
simple situation where we present the details of the calculations.

Next we consider the quantum system coupled to a general 
weak dissipative  medium. We take the case $\frac{k}{\omega} \sim 1$
when the effects on the dynamics due to the slow evolution and
the dissipative attenuation are of the same order. From eq. (\ref{41.2})
we obtain that the adiabatic nature of the phenomena disappears
if the non-unitary part of the Liouvillian  has non-null diagonal
terms in the instantaneous basis of Hamiltonian.

In the general case the loss of coherence due to the coupling to
a weak dissipative medium is a time-dependent phenomena
(see eqs. (\ref{43.3}) and (\ref{43.5})). In eq. (\ref{44.4}) we have
our main result where we obtain the condition that the coefficients
${\bar c}^{l_1 l_2}_{l_3 l_4} (t)$ have to satisfy in order to the imaginary
phase be geometric. This only happens when the interaction between 
the quantum system and its environment is time-dependent and is a
consequence of the time variation of a set of parameters. As in
the case of imaginary geometric phases appeared in the transition probabilities of
non-real Hamiltonians in the non-adiabatic regime\cite{joye,berry2,zwanziger}
here also the integral over the parameter space $\vec{\bf k}$ does not have to be a close
loop. The condition (\ref{44.4}) allows us to say directly from the expression
of ${\cal L}_D^H$ if the dissipative effect in the master equation
gives an imaginary geometric correction to the Berry phase\cite{berry}.
In the case where condition (\ref{44.4}) is fulfilled and the coefficients
$c^{ij}_{lm}(t)$ has an imaginary phase, the presence of dissipation gives
a correction to the real geometric phase.

The  condition (\ref{44.4})  is not satisfied by the coefficients
$ c^{l_1 l_2}_{l_3 l_4} (t)$ in the linear expansion  of the 
non-unitary part of the Liouvillian  (see eq. (\ref{40})) that represents 
the weak  interaction  of  a non-degenerate  quantum system  with a  reservoir
 at thermodynamic  equilibrium. This implies that for those type of couplings
 the imaginary phases are time-dependent.
In the present work we extend the validity of the   results derived for two
particular dissipation mechanisms discussed in reference \cite{JphysA}
that are rediscussed in section 4 using the approach presented in this work.

The non-hermitian parts in the Hamiltonians in references \cite{2,2.1,2.2} 
take into account the losses of
a quantum system to its environment 
(a suitable reservoir of degrees of freedom at equilibrium). 
In these models, 
the couplings between quantum systems and their neighborhoods in the
``{\it ab initio}'' Hamiltonians do not depend on the time variation of a
set of classical parameters. This fact, togheter with result (\ref{44.4}) 
(the condition to the existence of an imaginary geometric phase) 
put in check the correctness of
the imaginary geometric phases in the literature due to dissipative effects 
\cite{2,2.1,2.2}. 
In order to verify if the imaginary phase for a quantum model
described by a phenomenological non-hermitian hamiltonian truly
exists --- being of true geometric origin, and not a fake
one due to eqs. (75) and (76) --- one must apply the approach
presented here.
The presence of imaginary phases of geometric and time-dependent
natures in a quantum system yield distinct rates of loss of
coherence, and such distinction is experimentally
detectable.

Finally we point out that we have applied the results derived in here
 to quantum systems  in contact with  reservoirs
at thermodynamic equilibrium, but  they apply equally well to models whose interaction
with the environment  varies in time.  Only in this situation the imaginary
phase acquired by loss of coherence can be eventually geometric (if and
only if the condition (\ref{44.4}) is fulfilled).


\section*{\bf Acknowledgements}

A.C. Aguiar Pinto and M.T. Thomaz are in debt with E.V. Corr\^ea Silva
for the careful reading of the manuscript. A.C. Aguiar Pinto
thanks CNPq for financial support. M.T. Thomaz  thanks
CNPq for partial financial support.



\begin{thebibliography}{99}
\bibitem{nota} Like the case of no dissipation, the off-diagonal
elements of the density matrix are not uniquely defined unless we
impose a condition on $\langle u_i;t|\Big(\frac{d}{dt}|
u_i;t\rangle\Big)$, $i=1$ and $2$. In order to get eqs. (\ref{dom_45b})
and (\ref{dom_49b}), we
impose the conditions:  $\langle u_1;t|\Big(\frac{d}{dt}|
u_1;t\rangle \Big) = i\omega\sin^2(\frac{\theta}{2})$ and
$\langle u_2;t|\Big(\frac{d}{dt}| u_2 ;t\rangle\Big) = -
i\omega\sin^2(\frac{\theta}{2})$.
\end{thebibliography}
\end{document}